\documentclass[conference]{IEEEtran}
\IEEEoverridecommandlockouts
\usepackage{cite}
\usepackage{amsmath,amssymb,amsfonts}
\usepackage{algorithmic}
\usepackage{graphicx}
\usepackage{textcomp}
\usepackage{xcolor}
\def\BibTeX{{\rm B\kern-.05em{\sc i\kern-.025em b}\kern-.08em
    T\kern-.1667em\lower.7ex\hbox{E}\kern-.125emX}}
\begin{document}

\title{Poster: TM‑RugPull: A Temporally Sound Multimodal Dataset for Early Rug‑Pull Detection\\
{\footnotesize}
\thanks{}
}

\author{
\IEEEauthorblockN{Mohammad Pishdar}
\IEEEauthorblockA{
\textit{Department of Computer Science} \\
\textit{KU Leuven} \\
Leuven, Belgium \\
mohammad.pishdar@kuleuven.be
}
\and
\IEEEauthorblockN{Fatemeh Shoaei}
\IEEEauthorblockA{
\textit{Department of Computer Engineering} \\
\textit{Ilam University} \\
Ilam, Iran \\
f.shoaei@ilam.ac.ir
}
\and
\IEEEauthorblockN{Mozafar Bag-Mohammadi}
\IEEEauthorblockA{
\textit{Department of Computer Engineering} \\
\textit{Ilam University} \\
Ilam, Iran \\
m.bagmohammadi@ilam.ac.ir
}
\and
\IEEEauthorblockN{Mojtaba Karami}
\IEEEauthorblockA{
\textit{Department of Computer Engineering} \\
\textit{Ilam University} \\
Ilam, Iran \\
m.karami@ilam.ac.ir
}
\and
\IEEEauthorblockN{Bert Lagaisse}
\IEEEauthorblockA{
\textit{Department of Computer Science} \\
\textit{KU Leuven} \\
Leuven, Belgium \\
bert.lagaisse@kuleuven.be
}
}

\maketitle

\begin{abstract}
Rug pull is a critical attack in the world of blockchain technology. Despite this, the absence of sufficient time-bound and well-structured datasets is considered one of the significant issues faced while identifying early detection. Existing datasets do not provide the solution to this challenge because of temporal leakage or use of post-collapse indicators, insufficient modality coverage, and confusing or partial labels, especially with regards to DeFi tokens. To solve these problems, we present a highly curated and strictly time-bound dataset called TM-RugPull containing 1,000 projects, which include DeFi, meme, NFT, and celebrity token projects. We achieve temporal validation of the dataset by acquiring all three modalities, namely on-chain behavior, smart contract metadata, and OSINT signals. The project labels are provided based on manual investigation for entire projects’ lifespans and their collapse. Also, we make our dataset publicly available together with its codebase for data acquisition and feature extraction.
\end{abstract}


\section{Introduction}

Rug-pull scams have become one of the most devastating forms of fraud in
blockchain-based ecosystems, causing substantial financial losses to both retail
and institutional investors \cite{b1} \cite{b4}.
A rug pull is commonly defined as the abrupt removal of liquidity from, or
abandonment of, a token project by its creators \cite{b4}.
Despite the introduction of various detection approaches, the development of
techniques capable of identifying rug-pull events at an early stage remains
severely constrained by the nature and causality of existing datasets
\cite{b2} \cite{b3}.

Several limitations characterize existing publicly available rug-pull datasets.
Many suffer from temporal data leakage by relying on post-collapse signals, such
as liquidity drains, price crashes, which
leads to overestimated predictive performance and poor generalization in
real-world deployments \cite{b2}. Moreover, most datasets focus exclusively on
DeFi protocols, despite evidence that rug-pull behavior also affects meme
tokens, NFT-related projects, and celebrity or impersonation-based tokens
\cite{b4}.

Effective early-stage rug-pull detection therefore requires datasets that combine
temporal validity, multimodality, and operational relevance. On-chain behavioral
signals provide immutable evidence of token distribution and transactional
activity, while off-chain OSINT indicators—such as social media engagement and
web search trends—often precede on-chain anomalies \cite{b5}.
However, without strict temporal alignment, combining these modalities leads to
retrospective analysis and spurious inference \cite{b2}.

In order to tackle these issues, we propose TM-RugPull, a temporally consistent and multimodal dataset consisting of 1,000 token projects in various sectors. All blockchain and OSINT labels are assigned through a rigorous and expert-validated process. Our dataset contains no temporal leakage, goes beyond DeFi applications, and is accompanied by publicly accessible source code
\footnote{https://github.com/mohamadpishdar/RugPull-Defender/tree/main/src/DataSet}.

\section{DATASET DESIGN AND CONSTRUCTION}

Three key guiding principles underlie the development of TM-RugPull, which include temporal consistency, multimodal data, and manual labeling validation. The dataset is deliberately designed to facilitate the study of early-stage rug-pull events without any post-collapse information and incorporates a diverse set of token projects.

\subsection{Scope and Data Sources}
The TM-RugPull dataset consists of 1,000 genuine token projects deployed from 2016 to 2025 on five main blockchain networks, namely, Ethereum, Binance Smart Chain, Polygon, Arbitrum, and Fantom. In contrast to other datasets that are largely dominated by decentralized finance (DeFi) projects, the TM-RugPull dataset provides more comprehensive coverage by including different types of token projects, which include DeFi tokens, meme tokens, non-fungible token (NFT) and game projects, celebrity tokens, and impersonated tokens. Labels for each project in the TM-RugPull dataset are confirmed via cross-validation with publicly available scam/incident databases, such as those provided by CertiK, De.Fi, and Rekt.news, and then further validated by conservative manual analysis. For each project, data are acquired from three complementary channels, namely, on-chain behavioral features, smart contract metadata, and OSINT signals.

\subsection{Temporal Boundary and Leakage Prevention}
The midpoint is characterized by being the middle time between the deployment time of smart contracts and a timestamp determined uniquely for each project under investigation. The latter will not be the point of emergence of any potential red flags in the case of malicious projects but the point where the detection of the rug pull event occurs following a certain observation period within which the persistence of maliciousness is validated. Moreover, some exceptional operational events like contract upgrade/migration should be manually investigated to prevent false classification. Hence, all post-hoc actions like liquidity removal, transactions stopping, price drop, and social media activity are not included in the dataset. This ensures a purely causality-driven analysis of projects before the occurrence of the event being analyzed.
In terms of the generation of features, the TM-RugPull model uses both on-chain and off-chain OSINT features. In order to obtain on-chain features like the number of transactions, price changes, number of token holders, and token distribution, on-chain data must be manually analyzed using blockchain explorer tools like Etherscan and BscScan. This information is available on the contract’s dedicated page on blockchain explorers. Off-chain features are calculated based on OSINT signals through the estimation of temporal exposure of projects and their related posts on platforms like Google and Twitter, using search query techniques referred to as dorking.

\subsection{Labeling Protocol and Quality Contro}
The categorization of projects in TM-RugPull happens through a well-defined scientific and manual verification approach. A project will be categorized as a rug pull if and only if all the following characteristics are present in a 72-hour observation period: (i) almost complete absence of liquidity, (ii) lack of on-chain activity, and (iii) uncertainty about the price or practically no price and volume at all. The reason why a 72-hour observation period has been deliberately set is that legitimate operational events, such as disruptions and manually upgrading and replacing contracts, should already have been sorted out by this time to avoid any misclassification. Labels assigned are also independently verified through blockchain forensic reports and security incident records, alongside reaching consensus among two computer science experts.

\section{Dataset Overview \& Key Properties}

Table I provides a brief overview of TM-RugPull, including the temporal
coverage of the dataset, the types of blockchains covered, the token classes,
as well as the modalities of the data. TM-RugPull includes token projects
spanning the years 2016 to 2025 in different blockchain platforms and sectors.
TM-RugPull features a combination of multi-chain compatibility and multimodal
data collection through on-chain behavioral measures, metadata from smart
contracts, as well as temporally aligned OSINT attributes. 
All extracted attributes are solely dependent on the initial stage of each project’s life cycle
and are developed as candidate attributes for predictive modeling without any post hoc information.
Figure 1 shows the architecture of the data production process. On-chain and
OSINT attributes are aligned based on a unique project midpoint that sets the
upper time limit for generating the attributes, whereas independent forensic
evidence is only used for labeling.

\begin{figure*}[t]
    \centering
       \includegraphics[width=0.85\linewidth,height=5cm,keepaspectratio]{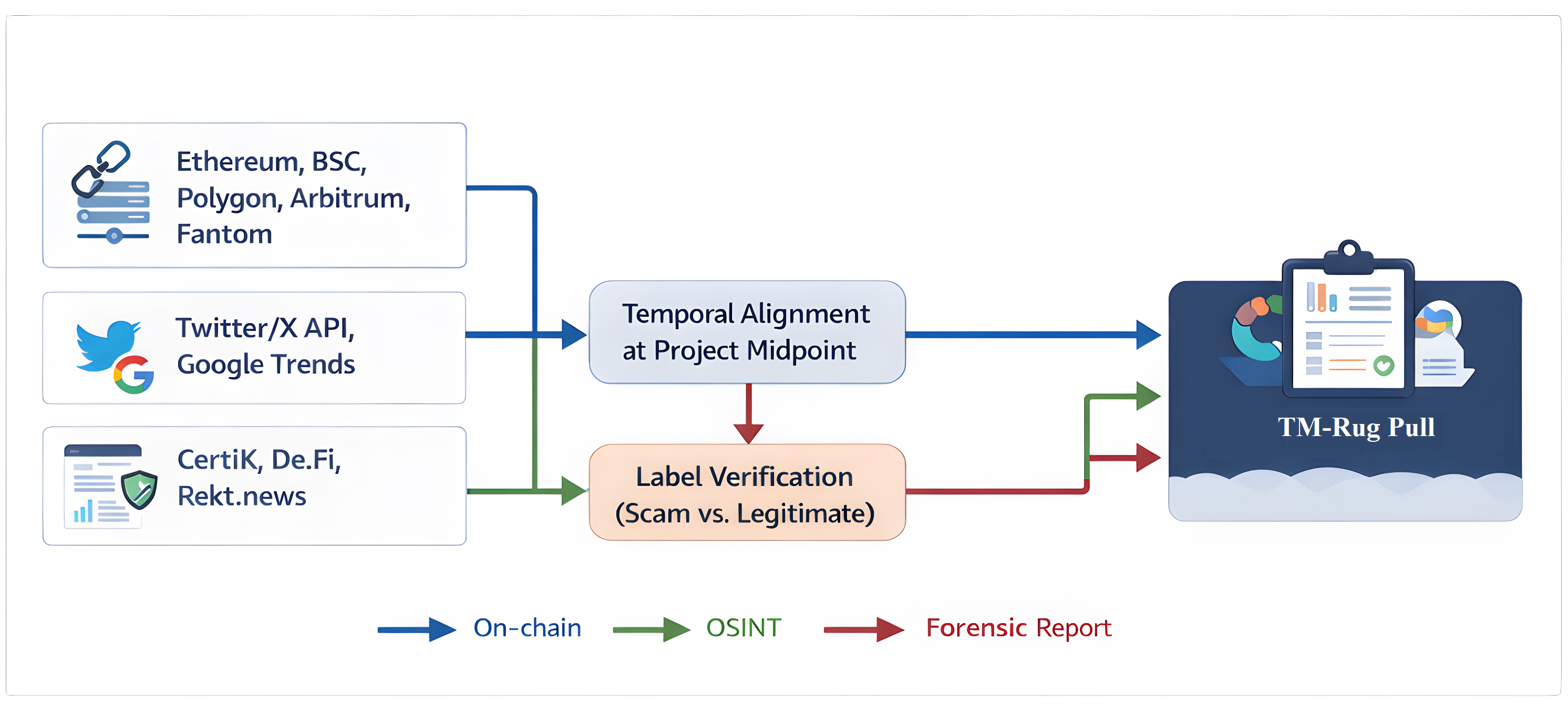}
    \caption{\small The pipeline for generating the dataset}
    \label{fig:data_pipeline}
\end{figure*}

\begin{table}[t]
\caption{Overview of TM-RugPull Dataset Composition}
\label{tab:dataset_composition}
\centering
\small
\setlength{\tabcolsep}{5pt}
\begin{tabular}{p{0.36\columnwidth} p{0.58\columnwidth}}
\hline
\textbf{Aspect} & \textbf{Description} \\
\hline
Dataset scope & 1,000 token projects collected between 2016 and 2025 \\
Token classes & Rug-Pull (59.9\%), Non-Rug-Pull (40.1\%) \\
Blockchains & ETH (63\%), BSC (32.1\%), Polygon (2.4\%), others (2.2\%) \\
Token categories & DeFi, Meme, NFT/Gaming, Celebrity \\
Data modalities & On-chain, Contract metadata, OSINT \\
Temporal scope & Pre-midpoint (first half of lifecycle) \\
Labeling & Forensic evidence \& expert consensus \\
\hline
\end{tabular}
\end{table}

\section{Motivation for TM‑RugPull}

A variety of open-source datasets have been developed to
study rug pull behavior within blockchain systems \cite{b1}.
Table II
presents a systematic comparison of such datasets according
to various attributes related to supervised learning, including
label existence, domain scope, presence of benign samples,
and temporal relevance.
From the point of view of supervised machine learning,
a dataset needs to fulfill three essential criteria. Firstly, it
should contain examples with explicit labels. Secondly, it
needs to contain examples for both malicious (rug-pull) and
non-malicious (non-rug-pull) behavior so that it can support
discriminative learning. Finally, it should have features which
represent behavioral signals rather than post-facto signals.
However, all of these three qualities do not exist simultaneously within any of the current rug-pull datasets.

Table II evaluates the current rug pull datasets, including Sun \cite{b7}, RPH (RPHunter) \cite{b12}, CRP (CRPWarner) \cite{b11}, Sol (SolRPDS for the Solana ecosystem) \cite{b8}, and TM, which refers to the proposed TM‑RugPull dataset.  based on criteria such as availability, blockchain and domain scope, inclusion of benign samples, temporality, multimodal data, and machine learning readiness (rows).

However, the TM-RugPull dataset has been specifically designed for machine learning research. The dataset includes well-labeled instances of projects with either a malicious or non-malicious classification; out of which around 59.9\% belong to the category of rug pull instances, while 40.1\% of the examples are benign instances. The benign examples included within the dataset are based on projects that have been chosen on the basis of their high security ratings from smart-contract security databases, including the CertiK database, and no incidents against them. Furthermore, the examples chosen undergo manual inspection prior to being included within the dataset. Additionally, the TM-RugPull dataset covers various domains apart from DeFi and includes time-aligned instances.

Table~\ref{tab:key_features} summarizes the descriptions of the main
columns included in the TM-RugPull dataset.

\begin{table}[t]
\caption{Comparison of Publicly Available Rug-Pull Datasets}
\label{tab:dataset_comparison}
\centering
\small
\setlength{\tabcolsep}{3pt} 
\begin{tabular}{p{0.22\columnwidth} @{}c c c c c@{}}
\hline
\textbf{Property} &
\textbf{Sun} &
\textbf{RPH} &
\textbf{CRP} &
\textbf{Sol} &
\textbf{TM} \\
\hline
Public &
\checkmark & \checkmark & \checkmark & \texttimes & \checkmark \\

Multi-Chain &
\checkmark & \texttimes & \texttimes & \checkmark & \checkmark \\

Multi-Domain &
\texttimes & \texttimes & \texttimes & \texttimes & \checkmark \\

Benign &
\texttimes & \checkmark & \texttimes & \checkmark & \checkmark \\

Early-Stage &
\texttimes & \texttimes & \texttimes & \texttimes & \checkmark \\

Multimodal &
\texttimes & \texttimes & \texttimes & \texttimes & \checkmark \\

ML-Ready &
\texttimes & \texttimes & \texttimes & \texttimes & \checkmark \\
\hline
\end{tabular}
\end{table}

\begin{table}[t]
\caption{Description of Key Features in the TM-RugPull Dataset}
\label{tab:key_features}
\centering
\small
\setlength{\tabcolsep}{5pt}
\begin{tabular}{p{0.33\columnwidth} p{0.62\columnwidth}}
\hline
\textbf{Feature} & \textbf{Description} \\
\hline
Blockchain & Target blockchain network where the token is deployed \\

Project start date & Timestamp corresponding to the first on-chain activity or smart contract deployment of the project. \\

Project end date & Timestamp marking the end of the observation window, used only for labeling and not for feature extraction. \\

Number of transactions & Total number of on-chain token transfer transactions. \\

Token holder count & Number of unique wallet addresses holding the token  \\

Token concentration ratio & Share of the total token supply controlled by top holders, measuring early ownership centralization. \\

Smart contract availability & Indicator showing whether verified smart contract source code is publicly available. \\

Online presence \& label &
Project visibility signals (X/Twitter and Google search presence) along
with the final Rug-Pull or Non-Rug-Pull label. \\
\hline
\end{tabular}
\end{table}

\end{document}